\documentclass[twocolumn]{svjour3}
\pdfpagewidth=\paperwidth
\pdfpageheight=\paperheight
\usepackage[scr=esstix,cal=boondox]{mathalfa} 
\usepackage[table]{xcolor}
\usepackage{enumitem}
\usepackage{graphicx}
\usepackage{verbatim}
\usepackage{color}
\usepackage{amsmath,amsfonts,amsbsy,amssymb}
\usepackage{upgreek}
\usepackage{natbib}

\usepackage{array}
\newcolumntype{L}{>{\centering\arraybackslash}m{3.5cm}}
\newcolumntype{P}{>{\centering\arraybackslash}m{4cm}}

\def\L{\mathcal L}

\usepackage{rotating}
\def\br{\boldsymbol {\rm r}}

\def\N{\mathcal N}
\def\r{{\boldsymbol {\mathcal r}}}

\def\P{{\mathcal P}}

\def\E{{\mathcal E}}

\def\r{\br}
\def\d{{\rm d}}

\def\D{{\mathcal D}}
\def\H{\mathcal H}
\def\Z{\mathcal Z}
\def\hr{{\hat \r}}
\def\B{{\mathcal B}}

\begin{document}\sloppy

\author{Paul N.\ Patrone${}^{1}$ \and Prajakta Bedekar${}^{1,2}$ \and  Nora Pisanic${}^{3}$ \and Yukari C.\ Manabe${}^{4}$ \and David L.\ Thomas${}^{4}$ \and Christopher D.\ Heaney${}^{3}$ \and Anthony J.\ Kearsley${}^{1}$}
\institute{1. National Institute of Standards and Technology \and 100 Bureau Drive, Gaithersburg, MD 20899, USA. \\ 2. Johns Hopkins University, Department of Applied Mathematics and Statistics, USA \\ 3. Johns Hopkins University, Bloomberg School of Public Health, USA \\ 4. Johns Hopkins University, School of Medicine, USA \\ \email{paul.patrone@nist.gov}
}

\title{Optimal Decision Theory for Diagnostic Testing: Minimizing Indeterminate Classes with Applications to Saliva-Based SARS-CoV-2 Antibody Assays}

\authorrunning{Minimizing Indeterminate Classes}
\titlerunning{Minimizing Indeterminate Classes}

\maketitle

\begin{abstract}{In diagnostic testing, establishing an indeterminate class is an effective way to identify samples that cannot be accurately classified.  However, such approaches also make testing less efficient and must be balanced against overall assay performance.  We address this problem by reformulating data classification in terms of a constrained optimization problem that (i) minimizes the probability of labeling samples as indeterminate while (ii) ensuring that the remaining ones are classified with an average target accuracy $X$.  We show that the solution to this problem is expressed in terms of a {\it bathtub principle} that holds out those samples with the lowest {\it local accuracy} up to an $X$-dependent threshold.  To illustrate the usefulness of this analysis, we apply it to a multiplex, saliva-based SARS-CoV-2 antibody assay and demonstrate up to a 30~\% reduction in the number of indeterminate samples relative to more traditional approaches.     }
\end{abstract}

\section{Introduction}

The SARS-CoV-2 pandemic has highlighted the importance of antibody testing as a means to monitor the spread of diseases such as COVID-19 (\cite{geography,EUA}).  But the widespread deployment of new assays has also revealed fundamental problems in the ability to analyze reliably the corresponding measurements.  Early on, this shortcoming was attributed to low prevalence, which made it difficult to distinguish true and false positives (\cite{challenges1}). However, it soon became clear that there were deeper issues related to statistical interpretation of raw data, suggesting the need to revisit the underlying theory of diagnostic classification (\cite{challenges2,Patrone2021,Chou21}). 

In this context, a fundamental problem arises when many measurements fall near a cutoff used to distinguish positive and negative samples.  The probability of correctly classifying these borderline cases hovers near 50~\%, so that even a small fraction thereof can significantly decrease overall accuracy.  A common solution is to define a third, {\it indeterminate} class for which one cannot draw meaningful conclusions, \textcolor{black}{although this is not always chosen to be near a cutoff (\cite{holdout1,holdout2,holdout3,holdout4,Nora1,Nora2,Heaney21})}.  While this approach increases the average accuracy for those samples that are classified, it also decreases testing efficiency.  Thus, there is a need to develop  strategies that balance the construction of indeterminate classes against overall assay performance.  

The present manuscript addresses this problem by answering the question: what classification scheme  (I) minimizes the fraction of indeterminate samples while (II) correctly identifying the remaining ones with a minimum average accuracy $X$?  When an indeterminate class is not permitted,\footnote{Reference (\cite{Patrone2021}) also defined a indeterminate class arising from an uncertain prevalence.  However, that task is distinct insofar as there is no constraint on the classification accuracy of the remaining samples.  Moreover, the adaptive prevalence-estimation algorithm in that work allows us to assume in this work that the prevalence is known.} common practice categorizes a sample as positive or negative if its measurement value $\br$ falls in a corresponding domain $D_P^\star$ or $D_N^\star$; see Fig.\ \ref{fig:rawdata}.  Moreover, it was recently shown that these domains can be optimized by solving an unconstrained optimization problem that maximizes accuracy associated with ``binary'' classification (\cite{Patrone2021}).  In contrast, the present work views (I) and (II) as a {\it constrained optimization} problem, with the size of the indeterminate class being the objective and the desired accuracy recast as a constraint.  We show that the solution to this problem extends the binary classification result by constructing the smallest indeterminate class via a ``bathtub principle'' applied to $D_P^\star$ and $D_N^\star$: one removes from them the measurements with the lowest probability of being correctly classified up to an $X$-dependent threshold.  As a practical matter, this ``waterline'' bounding the indeterminate domain can be efficiently and accurately estimated via numerical techniques such as bisection, making our result computationally tractable.  We provide examples and numerical validation using a saliva-based, multiplex SARS-CoV-2 antibody test, as well as mathematical proofs of our main results in the Appendix.

At the outset and in contrast with traditional methods, it is important to note that concepts such as specificity and sensitivity {\it per se} are not fundamental quantities of interest in our analysis.  As discussed in Sec.\ \ref{sec:discussion}, they describe the accuracy of a fixed classification scheme in two degenerate cases: 0~\% and 100~\% prevalence.  As such, it is trivial (but useless) to optimize either quantity by assigning all samples to a single class.  {\it Rather, we demonstrate that it is more useful to define accuracy as a prevalence-weighted, convex combination of specificity and sensitivity, since this naturally interpolates between the aforementioned degenerate cases.}  This choice also highlights an important (but often-ignored) fact: optimal classification domains, sensitivity, and specificity all change with prevalence.  {\it Thus, they are not static metrics of the assay performance in a setting where a disease is actively spreading.}  For more in-depth discussion, we refer the reader to Ref.\ (\cite{Patrone2021}), as well as Sec.\ \ref{sec:discussion} of the present manuscript.



We also emphasize that the concept of classification accuracy has both a {\it local} and {\it global} interpretation, and the interplay between these interpretations is fundamental to our analysis.\footnote{The testing community has largely restricted its attention to global assay properties, since regulatory reporting focuses on assay performance for large populations \cite{EUA}.}  In particular, one can construct conditional probability density functions (PDFs) $P(\r)$ and $N(\r)$ of a measurement outcome $\r$ -- i.e.\ a local property -- for (known) positive and negative samples.  As shown in Ref.\ (\cite{Patrone2021}), these PDFs are necessary to maximize the global  accuracy $X$, since the equation  
\begin{align}
pP(\r) = (1-p)N(\r) \label{eq:boundary}
\end{align}
defines the boundary between $D_P^\star$ and $D_N^\star$ when $p$ is the prevalence.  In the present work, we recast this observation by showing that $P(\r)$ and $N(\r)$ also directly define the local accuracy $Z(\r)$, and that its global counterpart $X$ is the average value of $Z(\r)$.  We next observe that the boundary given by Eq.\ \eqref{eq:boundary} is the set for which $Z=50$~\%, its lowest possible value.  The corresponding points are the first to be held out, since they contribute most to the average error.\footnote{An interesting corollary of the proofs in Ref.\ (\cite{Patrone2021}) is that $Z\ge 50$~\% for optimally defined classification domains without indeterminates.  Thus, we never need consider relative errors less than 50~\%.  See also  Sec.\ \ref{sec:theory} and the Appendix.}  Moreover, one sees that systematically removing the least accurate $\r$ yields the fastest increase in the global accuracy for the remaining points.  The bathtub principle formalizes this idea.


From a practical standpoint, the main inputs to our analysis are \textcolor{black}{training data associated with positive and negative samples; thus our approach is compatible with virtually any antibody assay.}  These data are used to construct the conditional PDFs $P(\r)$ and $N(\r)$, so that the classification and holdout problems are reduced to mathematical modeling.  This is also the key limitation of our approach insofar as such models are necessarily subjective.  However, this problem is not unique to our method.  Where possible, we incorporate objective information about the measurement process. See Sec.\ \ref{sec:example} and Ref.\ (\cite{Patrone2021}) for a deeper discussion of such issues and other limitations.


The remainder of this manuscript is organized as follows.  Section \ref{sec:notation} reviews key notation and terminology.  Section \ref{sec:theory} presents the general theory for defining optimal indeterminate domains.  Section \ref{sec:example} illustrates this analysis in the context of a saliva-based, multiplex SARS-CoV-2 saliva assay.  Section \ref{sec:validation} considers numerical validation of our analysis, and Section \ref{sec:discussion} concludes with a discussion and comparison with past works.  The Appendix provides a proof of our main result and other supporting information.  

\section{Notation and Terminology}
\label{sec:notation}

Our analysis is grounded in measure theory and set theory.  We review relevant concepts here. Readers well-versed in these topics may skip this section.  

\begin{figure}
\includegraphics[width=8cm]{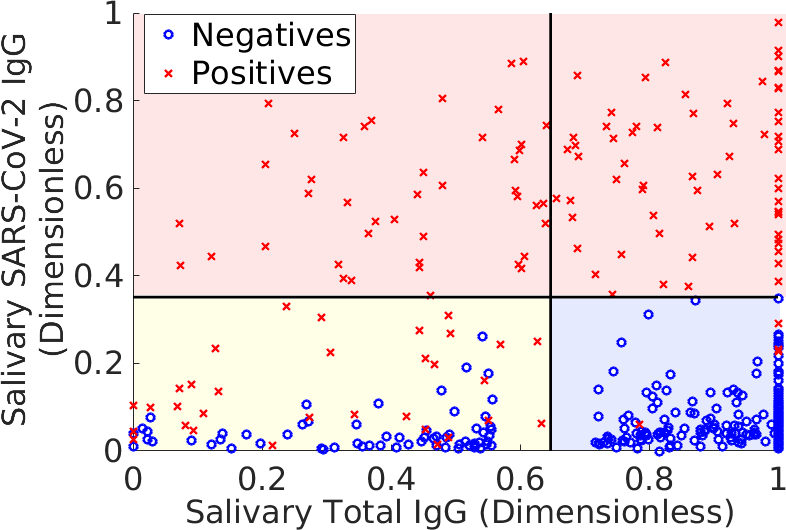}\caption{Training data associated with the Saliva assay described in Refs.\ (\cite{Nora1,Nora2}).  Red \textcolor{red}{x} denote known positives (confirmed via polymerase chain-reaction measurements), and blue \textcolor{blue}{o} denote pre-pandemic samples, which are assumed to be negative for SARS-CoV-2 antibodies.  The bold, horizontal and vertical black lines are cutoffs used to classify samples.  Data falling above the horizontal line (red shaded domain) are classified positive; data in the lower right box (shaded blue) are negative, and data in the lower left box (shaded yellow) are indeterminates.  The SARS-CoV-2 IgG measurements (vertical axis) are a sum of seven antibody levels measured by the assay, whereas the total IgG measurement (horizontal axis) is the total immunoglobulin-G (IgG) measurement as determined by an enzyme-linked immunosorbent assay (ELISA).}\label{fig:rawdata}
\end{figure}

\begin{itemize}
\item By a set, we mean a collection of objects, e.g.\ measurements or measurement values.  By a domain, we typically mean a set in some continuous measurement space; see, e.g.,\ Fig.\ \ref{fig:rawdata}.
\item The symbol $\in$ indicates set inclusion.  That is, $\br \in A$ means that $\br$ is in set $A$.
\item The symbol $\emptyset$ denotes the empty set, which has no elements.
\item The operator $\cup$ denotes the union of two sets.  That is, $C=A\cup B$ is  the set containing all elements that appear in either $A$ or $B$.
\item The operator $\cap$ denotes the intersection of two sets.  That is, $C=A\cap B$ is the set of elements shared by both $A$ and $B$.
\item The operator $/$ denotes the set difference.  We write $C=A/B$ to mean the set of all objects in $A$ that are not also in $B$.  Note that in general, $A/B \ne B/A$.  \textcolor{black}{Equivalently, $A/B$ can be interpreted as the ``subtraction'' or removal from $A$ of the elements it shares in common with $B$.}
\item The notation $A=\{\br: * \}$ defines the set $A$ as the collection of $\br$ satisfying condition $*$.

\end{itemize}

\textcolor{black}{Unless otherwise specified, the ``size'' or measure of a set refers to the probability of a sample falling within that set, i.e.\ its probability mass.  By the same token, we generally avoid using size to describe the actual dimensions (in measurement space) of a domain.}
Throughout we also distinguish between training data and test data.  The former is used to construct probability models, whereas the latter is the object to which the resulting classification test is applied.

\section{Minimum Probability Indeterminate Class}
\label{sec:theory}

We begin with the mathematical setting underlying classification.  Consider an antibody measurement $\r$, which can be a vector associated with multiple distinct antibody targets.  We take the set of all admissible measurements to be $\Omega$.  Our goal is to define three domains, $\D_P$, $\D_N$, and $\D_h$ associated with positive, negative, and indeterminate (or $h$ for ``hold-out'') samples.  In particular, we say that a {\it test sample} $\r$ is positive if it falls inside $\D_P$ (i.e.\ $\r\in \D_P$), and likewise for the other domains. 

We require that these domains have several basic properties to ensure that they define a valid classification scheme.  Recalling that $P(\r)$ and $N(\r)$ are conditional probabilities associated with positive and negative samples, define the measures of a set $S\subset \Omega$ with respect to $P$ and $N$ to be
\begin{subequations}
\begin{align}
\mu_P(S) &= \int_S \d \r P(\r) \label{eq:mup} \\
\mu_N(S) &= \int_S \d \r N(\r). \label{eq:mun}
\end{align}
\end{subequations}
That is, $\mu_P(S)$ is the probability of a positive sample falling in $S$, etc.  We then require that
\begin{align}
\mu_P(\D_P \cup \D_N \cup \D_h) = \mu_N(\D_P \cup \D_N \cup \D_h) = 1 \label{eq:normalization}
\end{align}
and 
\begin{align}
\mu_P(S \cap S') = \mu_N(S \cap S') = 0 \label{eq:disjoint}
\end{align}
when $S \ne S'$, for $S,S'$ chosen from $\D_P$, $\D_N$, or $\D_h$.  Equation \eqref{eq:normalization} states that the probability of any event falling in the positive, negative, or indeterminate domains is one; i.e.\ any sample can be classified.  Equation \eqref{eq:disjoint} states that the probability of a sample falling in more than one domain is zero, i.e.\ a sample has a single classification.  

Within this context, we define the total error rate to be
\begin{align}
\E[\D_P,\D_N] = \int_{\D_P} \d \r (1-p)N(\r) + \int_{\D_N} \d \r pP(\r) \label{eq:notnormederror}
\end{align}
where $p$ is the prevalence.  [See Ref.\ (\cite{Patrone2021}) for an unbiased method to estimate $p$ without needing to classify.]  The terms on the right-hand side (RHS) are the rates of false positives and false negatives.  Importantly, indeterminates are not treated as errors in Eq.\ \eqref{eq:notnormederror}, and $\E$ so defined is {\it not} the error rate of the assay restricted to samples that fall only within $\D_P$ and $\D_N$.  The latter is defined as 
\begin{align}
\E_r[\D_P,\D_N] = \frac{1}{p\mu_P(\D) + (1-p)\mu_N(\D)}\E[\D_P,\D_N] \label{eq:normalized}
\end{align}
where $\D=\D_P\cup \D_N$ is the set of all samples not in the indeterminate region.  \textcolor{black}{Note that Eq.\ \eqref{eq:normalized} is a conditional expectation; i.e.\ it is the average error conditioned on the set of samples that can be classified.}

In Ref.\ (\cite{Patrone2021}) we showed that \textcolor{black}{when the set $\mathcal Z_{1/2}=\{r: pP(\r) = (1-p)N(\r)\}$ has measure zero} and $\D_h$ is the empty set,\footnote{$\E$ and $\E_r$ are equal when $\D_h$ is the empty set.  \textcolor{black}{Note also that one can measure $\mathcal Z_{1/2}$ with respect to either $P$ or $N$.  This is because the set $\mathcal Z_{1/2}$ by definition entails that $p\mu_P(\mathcal Z_{1/2}) = (1-p)\mu_N(\mathcal Z_{1/2})$.}} $\E_r$ is minimized by the binary classification scheme
\begin{subequations}
\begin{align}
D_P^\star &= \{\r: pP(\r) > (1-p)N(\r)\} \label{eq:oldposopt} \\
D_N^\star &= \{\r:(1-p)N(\r) > pP(\r) \} \label{eq:oldnegopt}
\end{align}
\end{subequations}
for a prevalence $p$.  While $D_P^\star$ and $D_N^\star$ are not the optimal sets for the problem at hand, they play a fundamental role in the analysis that follows.\footnote{We use non-caligraphic symbols to denote binary classification sets, while we reserve caligraphic symbols for sets used in the holdout problem. }  \textcolor{black}{We also note an important corollary that when the $\mathcal Z_{1/2}$ has non-zero measure, Eqs.\ \eqref{eq:oldposopt} and \eqref{eq:oldnegopt} are generalized to
\begin{subequations}
\begin{align}
D_P^\star &= \{\r: pP(\r) > (1-p)N(\r)\} \cup Z_p\label{eq:newposopt} \\
D_N^\star &= \{\r:(1-p)N(\r) > pP(\r) \} \cup Z_n\label{eq:newnegopt}
\end{align}
\end{subequations}
where $Z_p$ and $Z_n$ are an arbitrary partition of $\mathcal Z_{1/2}$.  The physical interpretation of this generalization is that any point having equal probability of being negative or positive can be assigned to either class without changing the error.  In practice, however, classification often reverts to Eqs.\ \eqref{eq:oldposopt} and \eqref{eq:oldnegopt} as $\mathcal Z_{1/2}$ has zero measure for many practical PDFs.}

In the present work, we assume that there is a desired average accuracy $X$ and that $\L = 1-\E_r[D_P^\star,D_N^\star]<X$ when all samples are classified. Our goal is to define a minimum probability indeterminate class $\D_h^\star$ and domains $\D_P^\star$ and $\D_N^\star$ for which $\L[\D_P^\star,\D_N^\star]=X$; \textcolor{black}{that is, we wish to hold out the fewest samples so that those  remaining are classified with the desired accuracy}.  Mathematically, we seek to minimize
\begin{align}
\H[\D_h] = \int_{\D_h} \d\r\, Q(\r), \label{eq:objective}
\end{align}
where $Q(\r) = pP(\r) + (1-p)N(\r)$ is the probability of a {\it test} sample taking a value $\r$, subject to the constraint that
\begin{align}
p\int_{\D_P} \d \r P(\r) + (1-p)\int_{\D_N} \d \r N(\r) = X \int_{\D} Q(\r) \label{eq:constraint}
\end{align}
for $\D=\D_P \cup \D_N$.

To solve this problem, it is useful to introduce several auxiliary concepts.  In particular, define the local accuracy of the  unconstrained (i.e.\ no indeterminate), binary classification to be
\begin{align}
Z(\r,D_P,D_N)= \begin{cases}
pP(\r)/Q(\r) & r\in D_P \\
(1-p)N(\r)/Q(\r) & r \in D_N
\end{cases}
\end{align}
where $D_P$ and $D_N$ cover the whole set $\Omega$ up to sets of measure zero; moreover, let $Z^\star(\r)=Z(\r,D_P^\star,D_N^\star)$ be the local accuracy of the optimal solution to the binary problem.  Then the solution to the constrained problem given by Eqs.\ \eqref{eq:objective} and \eqref{eq:constraint} is
\begin{subequations}
\begin{align}
\D_h^\star &= \{\r: Z^\star(\r) < Z_0(X) \}\cup \mathcal C(X) \label{eq:optindeterminate}\\
\D_P^\star &= D_P^\star / \D_h^\star \label{eq:optpos} \\
\D_N^\star &= D_N^\star / \D_h^\star \label{eq:optneg}
\end{align}
\end{subequations}
where $Z_0(X)$ is the solution to the equation
\begin{align}
\int_{\Omega / \left\{ \{\r: Z^\star(\r) < Z_0 \} \cup {\mathcal C} \right \}} \d \r \left[Z^\star(\r) - X \right]Q(\r) = 0, \label{eq:zdef}
\end{align}
for any set $\mathcal C\subset \{\r: Z^\star(\r) = Z_0\}$ satisfying Eq.\ \eqref{eq:zdef}.   Proof of this result, as well as the strict interpretation of $\mathcal C$ requires significant analysis of Eq.\ \eqref{eq:constraint} and is reserved for the Appendix.  Here we provide an intuitive interpretation and describe a straightforward algorithm for computing Eqs.\ \eqref{eq:optindeterminate}--\eqref{eq:optneg}.  

Equation \eqref{eq:optindeterminate} informs that the points to hold out from classification are those with the lowest local accuracy up to some threshold value $Z_0$, which depends on $X$.  Equations \eqref{eq:optpos} and \eqref{eq:optneg} then amount to the observations that the positive and negative domains are the same as in the unconstrained binary problem, except that we remove the corresponding points with low enough local accuracy.  Equation \eqref{eq:zdef} requires that the average local accuracy for the classification sets $\D_P^\star$ and $\D_N^\star$ be $X$.  By virtue of the fact that $\D_h = \Omega / \D$, this fixes the boundary of the indeterminate set.  That is, the upper bound $Z_0(X)$ on the indeterminate local accuracy is the lower bound on the accuracy for sets that can be classified.  \textcolor{black}{The $\mathcal C(X)$ is a bookkeeping artifact accounting for the situation in which the set of points with local accuracy $Z_0(X)$ has non-zero probability mass.  In this case, not all of these points need to be held out if doing so would make $\L$ greater than $X$.  The choice of which points to make indeterminate then becomes subjective as they all have the same local accuracy.  In practice (e.g.\ for smooth PDFs), $\mathcal C(X)$ is a set of measure zero with respect to $Q$, so that we can ignore it in Eq.\ \eqref{eq:optindeterminate}.}

From Eqs.\ \eqref{eq:optindeterminate}--\eqref{eq:zdef} it is clear that determining $Z_0(X)$ is the key step in defining the optimal classification domains.  Fortunately, the interpretation afforded by Eq.\ \eqref{eq:zdef} leads to a straightforward bisection method.  First note that $1/2 \le Z^\star(\r) \le 1 $.  Let $\zeta_0=3/4$ be an initial guess for the value of $Z_0(X)$, and let $\zeta_j$ be the $j$th update computed iteratively as follows.  For each $\zeta_j$ compute $\D_P(\zeta_j)$, $\D_N(\zeta_j)$, as well as the left-hand side (LHS) of Eq.\ \eqref{eq:zdef}, which we denote by $\mathcal I_j$.  If $\mathcal I_j > 0$, then set $\zeta_{j+1} = \zeta_j - 2^{-(j+3)}$; if $\mathcal I_j < 0$, set $\zeta_{j+1} = \zeta_j + 2^{-(j+3)}$.  If $|I_j| \le \epsilon_X$ for some user-defined tolerance $\epsilon_X$, or if $j$ reaches some maximum iteration number $M$, stop the algorithm.  In the former case, the classified samples will have an average accuracy $\L$ in the range $X-\epsilon_X \le \L \le X + \epsilon_X$.   In the latter case, $\zeta_j - \epsilon_Z \le Z_0(X)=\zeta_j + \epsilon_Z$, where $\epsilon_Z \le 2^{-M+3}$ is the error in the estimate of $Z_0(X)$.  For context, 20 iterations of this algorithm yields errors $\epsilon_Z$ on the order of 1 in $10^7$.  \textcolor{black}{In the second case, the existence of a non-trivial set $\mathcal C(X)$ can be deduced from the observation that $\mathcal I_j$ does not converge, but rather cycles between two well-separated values, depending on whether $\zeta$ is greater than or less than $Z_0(X)$.  In this case, the set $\mathcal C(X)$ can be defined arbitrarily but consistent with Eq.\ \eqref{eq:zdef} once $Z_0(X)$ is identified to sufficient accuracy.}

\section{Example Applied to a Salivary SARS-CoV-2 IgG Assay}
\label{sec:example}

\begin{figure}
\includegraphics[width=8cm]{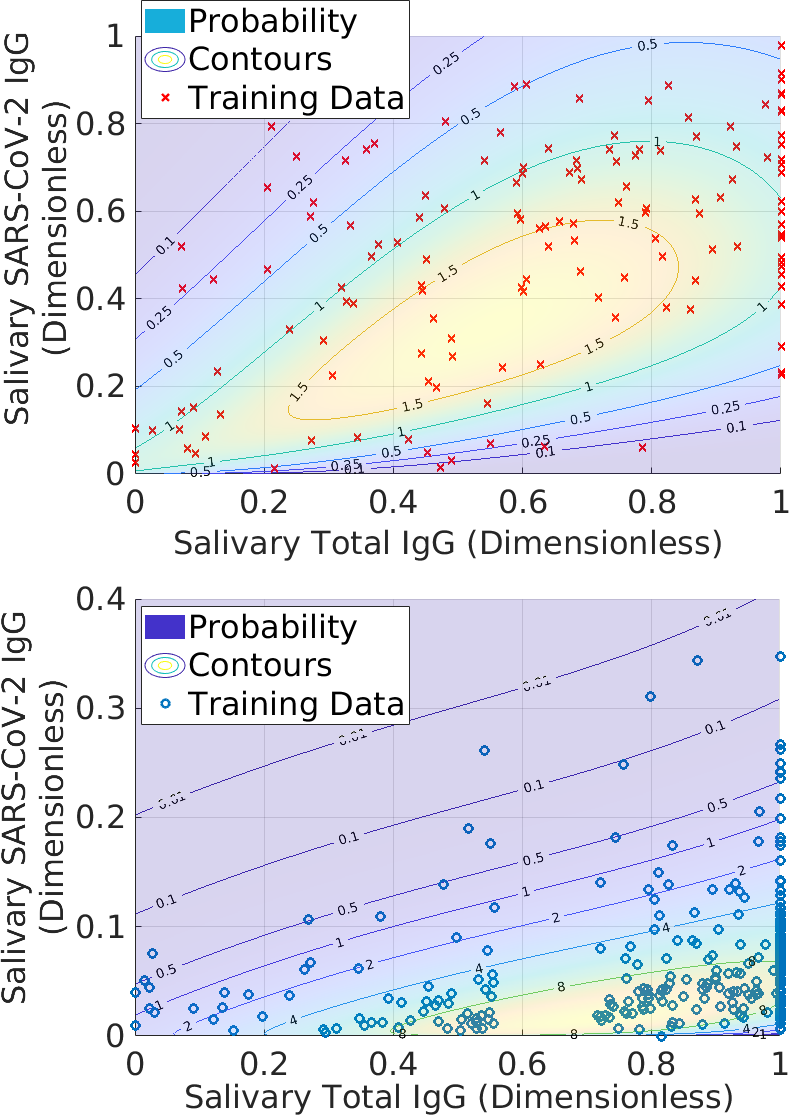}\caption{Probability density models associated with the training data.  See main text for a description of the probability density functions and the considerations behind their construction.  {\it Top}: Raw data and probability model for positive training samples.  {\it Bottom}: Negative training data and probability model.  }\label{fig:pdfs}
\end{figure}

To illustrate the analysis of Sec.\ \ref{sec:theory}, we consider a saliva-based assay described in Refs.\ (\cite{Nora1,Nora2}).  We refer the reader to those manuscripts for details of assay design, sample preparation, and measurement processes.  For each sample, two measurement values  are output: a total immunoglobulin G (IgG) enzyme linked immunosorbent assay (ELISA); and a sum of seven SARS-CoV-2 IgG measurements associated with distinct antigen targets.  As a preliminary remark, we observe that the numerical range of the data spans several decades of median fluorescence intensity (MFI), which is difficult to model directly.  We also note that the measurements are bounded from below by zero and have a finite upper bound.  This motivates us to transform each numerical value $d$ via $\log_2[d+2]-1$, which corresponds representing the data in terms of bits.  Empirically we also find that this transformation better separates positive and negative populations.  Total IgG values are then rescaled to the domain $[0,1]$ by dividing each measurement by the maximum.  SARS-CoV-2 measurements are similarly rescaled to the domain $[0,1]$, although we divide the log-transformed data by $7$, since there were no samples with saturated values. After transformation, each sample is represented by a two-dimensional vector $\r=(x,y)$, where $x$ is the normalized total IgG value, and $y$ is the normalized SARS-CoV-2 counterpart.

The results of this transformation are shown in Fig.\ \ref{fig:rawdata}, along with classification domains currently \textcolor{black}{used} with this assay.\footnote{All data correspond to samples for which more than 14 days have elapsed since symptoms onset.  Also, the original training total IgG data included samples that were diluted to achieve measurement values above a saturation value.  All such data were rounded down to the undiluted upper threshold to be consistent with the validation data.  This amounts to data censoring, for which we can still define the relevant likelihood functions used in parameter estimation.  See also the Appendix.}  The goal of the analysis is to maintain accuracy while decreasing the number of indeterminate samples by finding the domain $\D_h$ with the smallest {\it probability mass}.   \textcolor{black}{We remind the reader that size does not refer to the (generalized) volume in measurement space.  Rather it refers to the fraction of samples expected to fall within the domain, since this is what controls the number of indeterminate samples.  Thus, it is possible that $\D_h$ can be quite large when expressed in terms of antibody levels and yet contain very few samples.}

To motivate our probability models, we consider the phenomena that could affect measurements.  In particular, we anticipate that for positive samples, there should be a degree of correlation between total IgG and SARS-CoV-2 specific antibodies.  However, at extreme total IgG values, the SARS-CoV-2 levels may become independent as (i) all measurements will revert to noise when $x\to -\infty$ or (ii) SARS-CoV-2 antibody levels will decouple from total antibody levels when the latter is excessively high, e.g.\ if an individual has been exposed to a large number of different pathogens.  \textcolor{black}{We also recognize that the ELISA instrument only reports numerical values on the domain $[x_{\min},x_{\max}]$.  Thus, fluorescence levels above $x_{\max}$ are rounded down to the upper bound, and levels below $x_{\min}$ are rounded up to the lower bound}.  As shown in Fig.\ \ref{fig:rawdata}, this has the effect of accumulating data (and thus probability mass) on the lines $x=x_{\min}$ and $x=x_{\max}$.  While details are reserved for the Appendix, this observation leads us to model positive and negative samples via a PDF of the form 
\begin{align}
 P(x,y) &= \mathcal P_0(x,y) + \mathcal P_l(y)\delta (x) + \mathcal P_r(y)\delta(x-1), \label{eq:formalPDF}
\end{align}
where $0 \le x \le 1$, $0 \le y < 1$, $\delta(x)$ is the Dirac delta function, and $\mathcal P_0(x,y)$ is assumed to be bounded and continuous on the whole domain.  The functions $\mathcal P_l(y)$ and $\mathcal P_r(y)$ characterize the probability of SARS-CoV-2 antibody levels for measurement values saturated at the left ($l$) and right ($r$) bounds.  We emphasize that the use of delta functions in Eq.\ \eqref{eq:formalPDF} is formal and should be treated with care.  A more rigorous interpretation of what is meant by Eq.\ \eqref{eq:formalPDF} is discussed in the Appendix.

To model the function $\mathcal P_0(x,y)$, we treat the total IgG measurements as independent normal random variables with an unknown mean and variance.  {\it Within} the domain $0 < x < 1$ (note the strict inequalities) and $0\le y \le \infty$, we assume that the SARS-CoV-2 measurements are well described by a Gamma distribution with a fixed (but unknown) scale factor and shape parameter with a sigmoidal dependence on $x$.  This dependence is motivated by the correlation described previously.  Taken together, this yields the PDF
\begin{subequations}
\begin{align}
\mathcal P_0(x,y) &= \frac{e^{-(x-\mu)^2/(2\sigma^2)}}{\sqrt{2\pi}\sigma}y^{k(x)-1}\frac{e^{-y/\theta}}{\Gamma(k(x))\theta^{k(x)}} \\
k(x) &= \alpha_1^2 [{\rm \tanh}(\alpha_2 (x-\alpha_3))+1] + \alpha_4^2
\end{align}
\end{subequations}
where $\mu$, $\sigma$, $\theta$, and the $\alpha_j$ are to-be-determined.  The boundary functions are defined to be
\begin{align}
\mathcal P_l(y) &= \frac{y^{k(0)-1}e^{-y/\theta}}{\Gamma(k(0))\theta^{k(0)}} \int_{-\infty}^0 \d x \frac{e^{-(x-\mu)^2/(2\sigma^2)}}{\sqrt{2\pi}\sigma} \\
\mathcal P_r(y) &= \frac{y^{k(1)-1}e^{-y/\theta}}{\Gamma(k(1))\theta^{k(1)}} \int_{1}^{\infty} \d x \frac{e^{-(x-\mu)^2/(2\sigma^2)}}{\sqrt{2\pi}\sigma} 
\end{align}
which describes the probability that a total IgG value below (above) $x=0$ ($x=1$) will be mapped back to the lower (upper) instrument bound.  The free parameters are determined via maximum likelihood estimation using a censoring-based technique; see the Appendix.  As an approximation, we truncate the $y$-domain to be $0 \le y \le 1$ and renormalize the resulting PDF on this domain.  

For the negative PDF $N(x,y)$, we anticipate that non-specific binding of the total IgG antibodies to the SARS-CoV-2 antigens will lead to a degree of correlation, albeit to a less extent than for positives.  Thus, we use the same form of $P(x,y)$, but refit the parameters using the negative training data.    Figure \ref{fig:pdfs} shows the outcome of this exercise for the two training sets.  \textcolor{black}{Because $\mathcal P_l(y)$, $\mathcal P_r(y)$, and corresponding terms for $N(x,y)$ are continuous with respect to the Gamma portion of $P(x,y)$ and $N(x,y)$, the former can be inferred from the contour lines in the figure (up to a normalization factor) and are thus not shown.} 

\begin{figure}
\includegraphics[width=8cm]{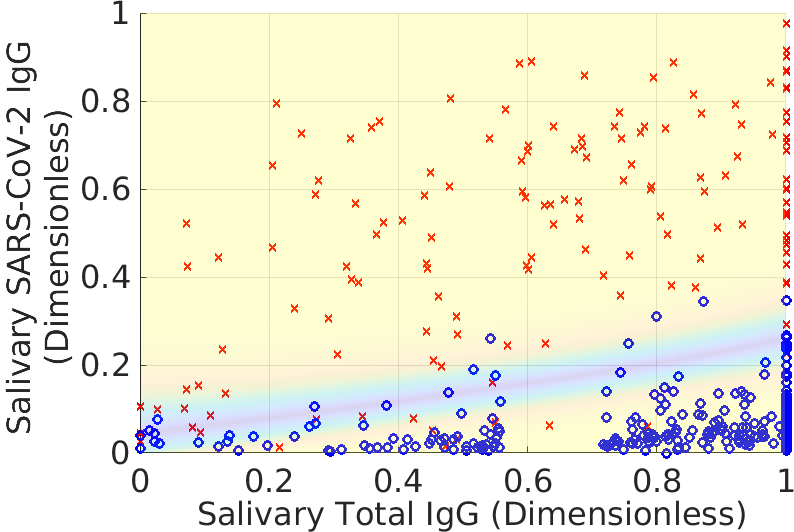}\caption{Local accuracy $Z^\star(\br)$ of the assay according to the probability models shown in Fig.\ \ref{fig:pdfs}.  Note that $Z^\star(\br)$ approaches 100~\% in regions where $P(\br)$ and $N(\br)$ do not overlap.  Conversely, in regions where the PDFs overlap, it is more challenging to correctly identify samples.  Thus $Z^\star(\br)$ decreases towards its minimal value of 1/2 in such regions.  Note that $Z^\star(\br)$ is never less than 1/2 (50/50 odds of correct classification).   
}\label{fig:relac}
\end{figure}

\begin{figure}
\includegraphics[width=8cm]{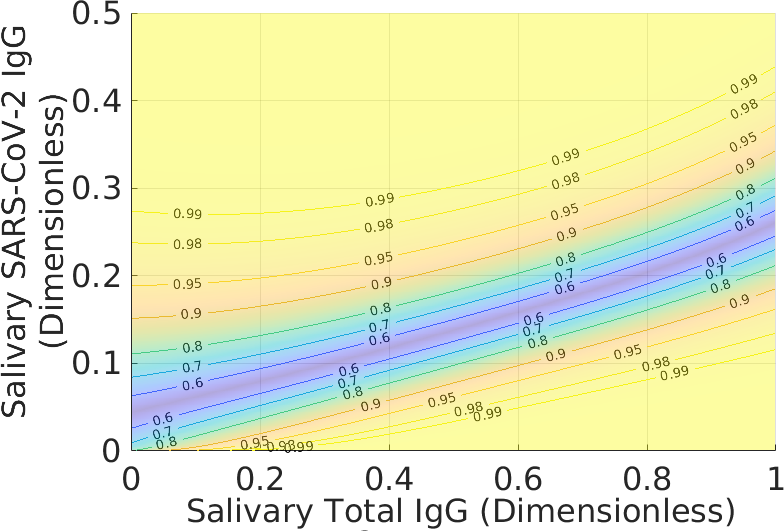}\caption{Illustration of the bathtub principle used to compute the minimal probability indeterminate domain.  The contour lines are different ``waterlines'' up to which we can hold out samples.  The label on each contour is the local accuracy of the assay.  In order to define the indeterminate region, we use the target {\it global} accuracy $X$ to define a maximum {\it local} accuracy up to which we hold out samples.  Increasing the global accuracy of the restricted classification increases the waterline, thereby holding out more samples.  }\label{fig:waterline}
\end{figure}

\begin{figure}
\includegraphics[width=8cm]{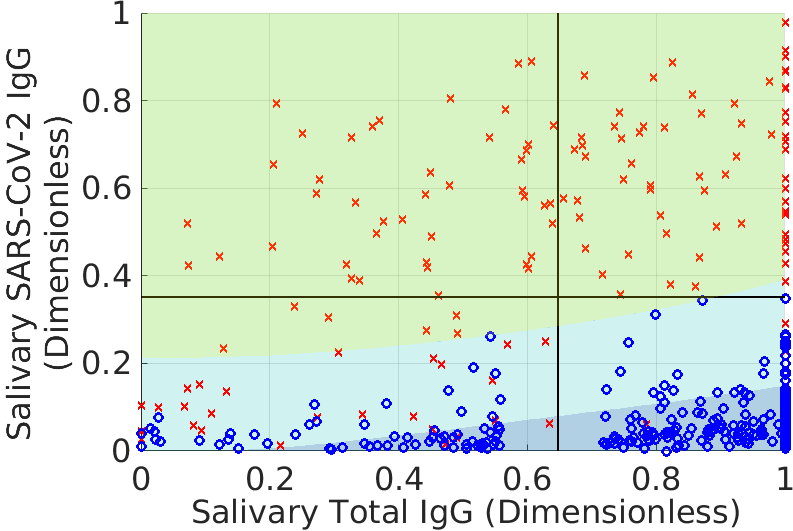}\caption{Positive (yellow-green), negative (dark blue), and indeterminate (light-blue) classification domains  defined for a theoretical target accuracy of $99.6$~\% for the training data in the previous figures.  Symbols have the same meaning as in previous figures. The empirical accuracy is 98.8~\%, with a specificity of 100~\% and sensitivity of 96.7~\%.  The total accuracy is the prevalence-weighted combination of these latter quantities.  Note the prevalence is associated with the restricted set of samples that are actually classified; see Sec.\ \ref{sec:discussion}.  Discrepancy between the theoretical and empirical accuracies is due to idealization of the modeling and stochasticity in the data.  For comparison, the horizontal and vertical black lines are the same as in Fig.\ \ref{fig:rawdata} and denote the corresponding classification domains originally used for this assay.  The indeterminate region based on the bathtub principle reduces the number of unclassified samples by more than 12~\% relative to the original domains while maintaining specificity and improving sensitivity for the training data.  See also Table \ref{tab1} and Sec.\ \ref{sec:discussion} for other examples of indeterminate domains.}\label{fig:holdout}
\end{figure}

\begin{figure}
\includegraphics[width=8cm]{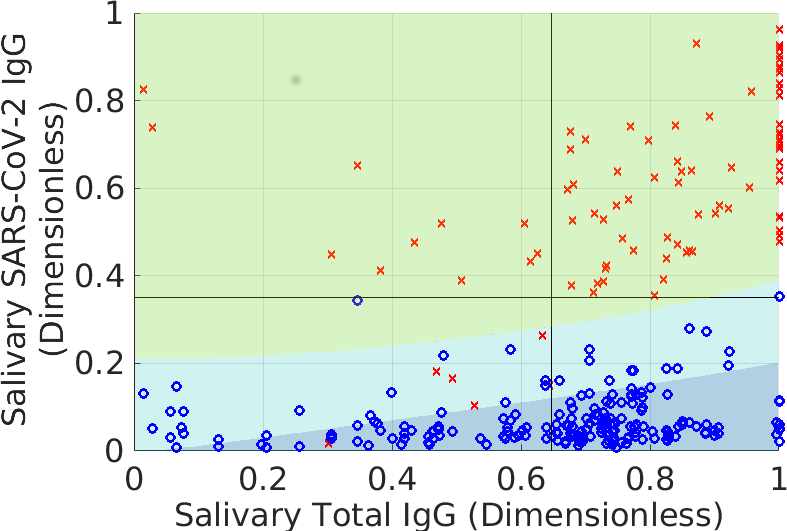}\caption{Positive (yellow-green), negative (dark blue), and indeterminate (light-blue) classification domains for validation data and defined for a theoretical target accuracy of $99.6$~\%.  The validation data was not used for training the probability models.  Symbols have the same meaning as in previous figures. The empirical accuracy is 99.2~\%, with a specificity of 99.4~\% and sensitivity of 98.8~\%.  The indeterminate region based on the bathtub principle reduces the number of unclassified samples by almost 40~\%.  See Sec.\ \ref{sec:discussion} for other examples of indeterminate domains.}\label{fig:validation}
\end{figure}

Figures \ref{fig:relac} and \ref{fig:waterline} show $Z^\star(\br)$ and waterlines necessary to achieve different average accuracies.  The bathtub principle is  shown in the latter; see also Ref.\ (\cite{bathtub}) for related ideas.  To ensure that $\L=X$, we only hold out samples up to the corresponding value of $Z_0(X)$.  Note that indeterminates are concentrated in regions where there is significant overlap between positive and negative samples.  Figure \ref{fig:holdout} shows the corresponding classification domains computed according to the bathtub principle for a target accuracy of $99.6$~\%; see also Table \ref{tab1}.  Relative to the original classification domains, the analysis reduces the empirical rate of indeterminate samples by more than 11~\% while increasing both accuracy and sensitivity of the assay (with empirical specificity remaining constant).  See also Fig.\ \ref{fig:validation} and Sec.\ \ref{sec:discussion} for additional examples of holdout domains.  

\begin{table*}
\begin{tabular}{ | l | L  P | L | }\hline
   {\bf Data \& Method} & {\bf COVID Samples}  & {\bf Pre-COVID Samples} & {\bf All Samples}  \\ \hline
  {\bf Training Samples } & 147 & 283 & 430 \\
 \rowcolor{yellow!15} \qquad Holdouts (Rectilinear)  & 32/147, 21.8~\% & 64/283, 22.6~\% & 96/430, 22.3~\% \\
  \qquad \cellcolor{blue!15}Holdouts (Optimal)  &\cellcolor{blue!15} 28/147, 19.1~\%  & \cellcolor{blue!15}56/283, 19.8~\% &\cellcolor{blue!15} 84/430, 19.5~\% \\ &&&  \\
  & {\bf Sensitivity} & {\bf Specificity} & {\bf Accuracy} \\
  \qquad \cellcolor{yellow!15}Classification  (Rectilinear) &\cellcolor{yellow!15} 111/115, 96.5~\% \qquad   [92.0~\%, 98.9~\%] &\cellcolor{yellow!15} 219/219, 100~\% \qquad\qquad [98.6~\%, 100~\%] &\cellcolor{yellow!15} 330/334, 98.8~\% \qquad [97.2~\%, 99.6~\%] \\ 
  \qquad \cellcolor{blue!15}Classification  (Optimal) & \cellcolor{blue!15}115/119, 96.6~\% \qquad [92.3~\%, 99.0~\%]& \cellcolor{blue!15}227/227, 100~\% \qquad \qquad [98.7~\%, 100~\%] & \cellcolor{blue!15}342/346, 98.8~\% \qquad [97.3~\%, 99.6~\%]\\ \hline \hline
  {\bf Validation Samples} & 87 & 192 & 279 \\
  \qquad \cellcolor{yellow!15}Holdouts (Rectilinear) & \cellcolor{yellow!15}6/87, 6.9~\% & \cellcolor{yellow!15}66/192, 34.4~\% & \cellcolor{yellow!15}72/279, 25.8~\% \\
    \qquad \cellcolor{blue!15}Holdouts (Optimal)  & \cellcolor{blue!15}5/87, 5.8~\% & \cellcolor{blue!15}34/192, 17.7~\% & \cellcolor{blue!15}39/279, 14.0~\% \\ &&&
\\  
  & {\bf Sensitivity} & {\bf Specificity} & {\bf Accuracy} \\    
    
  \qquad \cellcolor{yellow!15}Classification  (Rectilinear) & \cellcolor{yellow!15}81/81, 100~\% \qquad \qquad [96.3~\%, 100~\%] & \cellcolor{yellow!15}125/126, 99.2~\% \qquad \qquad [96.3~\%, 100~\%] & \cellcolor{yellow!15}206/207, 99.5~\% \qquad [97.7~\%, 100~\%] \\ 
  \qquad \cellcolor{blue!15}Classification  (Optimal) & \cellcolor{blue!15}81/82, 98.8~\% \qquad \qquad [94.4~\%, 100.0~\%] & \cellcolor{blue!15}157/158, 99.4~\% \qquad \qquad [97.0~\%, 100~\%] & \cellcolor{blue!15}238/240, 99.2~\% \qquad [97.3~\%, 99.9~\%] \\ \hline 
\end{tabular}\caption{Summary of fraction of holdouts, sensitivity, and specificity for the data in Figs.\ \ref{fig:holdout} and \ref{fig:validation}.    The rectilinear classification method is described in Fig.\ \ref{fig:rawdata}, while the optimal method is given by Eqs.\ \eqref{eq:optindeterminate}--\eqref{eq:zdef}. For sensitivity, specificity, and accuracy calculations, the numbers in brackets are empirical 95~\% confidence intervals.}\label{tab1}
\end{table*}

\section{Numerical Validation}
\label{sec:validation}

To validate that the sets $\D_P^\star$, $\D_N^\star$, and $\D_h^\star$ obtained in Sec.\ \ref{sec:theory} are optimal, we consider a numerical experiment wherein we perturb $\H$ as a function of these domains.    For point $\br\in \D_h$ and $\r'\in \D$, we {\it formally} define a ``point-swap derivative'' to be
\begin{align}
\frac{\delta \H[\D_h]}{\delta \br \delta \br'} = \frac{Z(\r)-X}{Z(\r') - X}.  \label{eq:setderivative}
\end{align} 
In principle $Z(\r)$ can be an arbitrary definition of local accuracy, although in practice we take $Z(\r) = Z^\star(\r)$ in this section.  The interpretation of Eq.\ \eqref{eq:setderivative} is as follows.  In taking point $\r'$ from $\D$ and adding it to $\D_h$ and vice-versa for  $\r$, we must ensure that the constraint Eq.\ \eqref{eq:constraint} remains satisfied.  The ratio $\frac{Z(\r)-X}{Z(\r') - X}$ provides the ``rate-of-exchange'' of probability.  For example, if $Z(\r)-X < Z(\r') - X < 0$, then adding $\r$ to $\D$ will {\it infinitesimally} decrease the global accuracy, so that we must hold out a larger yet still infinitesimal fraction of $Q$ in the vicinity of $\r'$.  It is clear that Eq.\ \eqref{eq:setderivative} goes through a singularity when $Z(\r')\to X$ and becomes negative for $Z(\r')>X$ and $Z(\r) < X$.  The interpretation of this is straightforward: we should always reverse any swap for which a point with local accuracy greater than the average is put in the indeterminate class.  Such points are not considered in the analysis below.  More rigorous interpretations of Eq.\ \eqref{eq:setderivative} are considered in the appendix, especially in the context of the singular PDF given by Eq.\ \eqref{eq:formalPDF}.

\begin{figure}
\includegraphics[width=8cm]{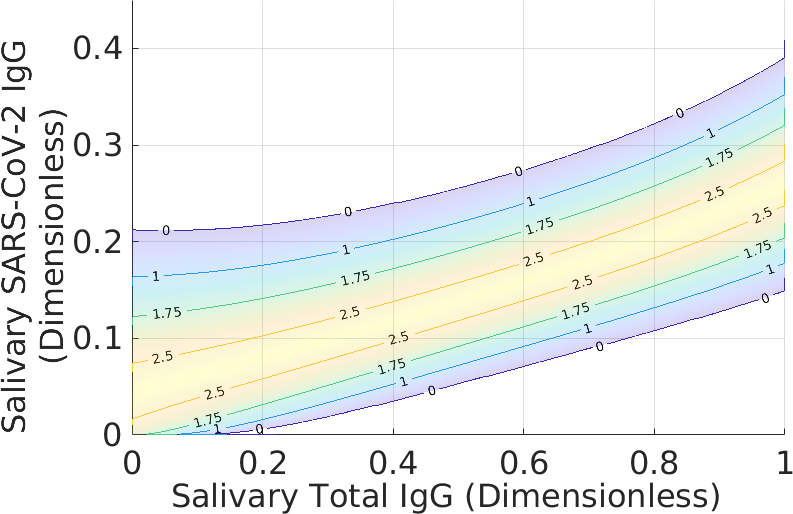}\caption{The logarithm of the swap derivative given by Eq.\ \eqref{eq:setderivative} computed for the optimal domains $\D^\star_P$, $\D^\star_N$, and $\D^\star_h$.  The 0-level line is the boundary of the indeterminate region.  Note that the logarithm is everywhere positive.  Thus, swapping any infinitesimal regions between $\D^\star_h$ and $\D^\star$ will increase the probability mass in the indeterminate, provided constraint Eq.\ \eqref{eq:constraint} is satisfied.}\label{fig:swapderiv}
\end{figure}

The benefit of Eq.\ \eqref{eq:setderivative} is that it allows us to estimate a ``set-partial derivative'' by computing the relative probability exchange for any point in the indeterminate domain.  In particular, we compute
\begin{align}
\frac{\delta \H[\D_h]}{\delta \br } = \inf_{\substack{\r' \\ Z(\r') < X \\ \r' \in \D}} \left[\frac{Z(\r)-X}{Z(\r') - X} \right] \label{eq:minswap}
\end{align}  
for the optimal domains $\D_h^\star$ and $\D^\star$.  Figure \ref{fig:swapderiv} shows the logarithm of Eq.\ \eqref{eq:minswap} for a mesh of points in the indeterminate region, taking $Z(\r) = Z^\star(\r)$.  Note that swapping any point in the indeterminate region with one in the positive and negative classification domains increases the size of the indeterminate, as expected.  

To validate that swapping points between $\D_P^\star$ and $\D^\star_N$ does not increase the accuracy of the assay or decrease the size of the indeterminate domain, we examine the quantity $Z(\r)$ directly.  In particular, the Appendix shows that $Z^\star(\r) \ge 1/2$ for all $\r \in \D^\star$ guarantees that $\D_P^\star=D^\star_P / \D_h^\star$ and $\D_N^\star=D_N^\star / \D_h^\star$ are optimal for the indeterminate region $\D_h^\star$.  Figure \ref{fig:relac} demonstrates that this inequality holds for the solution given by Eqs.\ \eqref{eq:optindeterminate}--\eqref{eq:zdef}.  Thus, no rearrangement of points decreases the size of the indeterminate domain.

\section{Discussion: Historical Context, Open Direction, and Limitations}
\label{sec:discussion}


\subsection{The Role of Prevalence}
\label{subsec:prev}

Examination of Eq.\ \eqref{eq:constraint} reveals that the terms of the LHS are proportional to prevalence-weighted estimates of sensitivity and specificity.  In particular, recognize that
\begin{subequations}
\begin{align}
S_e &= \left[ \int_{\D}P(\r) \, \d\r \right]^{-1} \int_{\D_P} P(\r) \, \d\r, \label{eq:sensitivity}\\
S_p &= \left[ \int_{\D}N(\r) \, \d\r \right]^{-1} \int_{\D_N} N(\r) \, \d\r \label{eq:specificity}
\end{align}
\end{subequations}
are the sensitivity and specificity restricted to the domain $\D$.  When there is no indeterminate domain, the normalization factors $\int_{\D} P(\r) \d\r= \int_{\D} N(\r)\d\r=1$, so that Eqs.\ \eqref{eq:sensitivity} and \eqref{eq:specificity} revert to the standard definitions of these quantities.  In this case, we see that Eq.\ \eqref{eq:constraint}, which no longer acts as a constraint, amounts to the statement that the prevalence-weighted sum of sensitivity and specificity is equal to $X$; that is
\begin{align}
pS_e + (1-p)S_p = X. \label{eq:noholdacc}
\end{align}

When we permit an indeterminate class, however, the interpretation is not as straightforward.  In particular, the presence of the term $\mathscr N_Q=\int_\D Q(\r) \d\r$ on the right-hand side (RHS) appears problematic, for note that it implies
\begin{align}
\mathscr N_Q^{-1}\left[ p \int_{\D_P} P(\r) \d\r + (1-p)\int_{\D_N} N(\r)\d\r \right ] = X. \label{eq:conundrum}
\end{align}
The normalization factor $\mathscr N_Q$ differs from its counterparts in Eqs.\ \eqref{eq:sensitivity} and \eqref{eq:specificity}.  Thus, it is not obvious what our constraint enforces about the sensitivity and specificity of the assay restricted to $\D$.

The resolution to this conundrum is to recognize that the {\it prevalence of the population also changes when we restrict classification to $\D$}.  This is not to say that the value of $p$ itself (i.e.\ associated with the total population) changes, but rather that the relative fraction of positives and negatives differs on $\D \subset \Omega$.  This is not unexpected, since the shape of the indeterminate region is a function of the local accuracy $Z$, which depends on the specifics of the probability models.  Mathematically, we understand these observations by rewriting Eq.\ \eqref{eq:conundrum} in the form
\begin{align}
\frac{p\mathscr N_P}{\mathscr N_Q} \int_{\D_P} \frac{P(\r)}{\mathscr N_P} \d\r + \frac{(1-p)\mathscr N_N}{\mathscr N_Q} \int_{\D_N} \frac{N(\r)}{\mathscr N_N} = X \label{eq:solconundrum}
\end{align}
where $\mathscr N_P = \int_{\D} P(\r) \d\r$ and $\mathscr N_N = \int_{\D} N(\r) \d\r$ are the required normalization constants.  Equation \eqref{eq:solconundrum} becomes an analogue to Eq.\ \eqref{eq:noholdacc} of the form
\begin{align}
\frac{p\mathscr N_P}{\mathscr N_Q} S_e + \frac{(1-p)\mathscr N_N}{\mathscr N_Q} S_p = X \label{eq:noconundrum}
\end{align}
where $p_\D = p\mathscr N_P/\mathscr N_Q$ is the prevalence restricted to the domain $\D$.  Note that $p_\D$ has the properties necessary to be a prevalence:
\begin{align}
\frac{p\mathscr N_P}{\mathscr N_Q} + \frac{(1-p)\mathscr N_N}{\mathscr N_Q} =1 \implies 1-p_\D =  \frac{(1-p)\mathscr N_N}{\mathscr N_Q} \label{eq:restrictedprev}
\end{align}
which is a consequence of the definition of $\mathscr N_Q$.  Thus, we see that the constraint corresponds to a domain-restricted-prevalence weighted sum of sensitivity and specificity.

From a theoretical standpoint, Eq.\ \eqref{eq:restrictedprev} is extremely serendipitous.  The constraint as defined by Eq.\ \eqref{eq:constraint} only refers to the prevalence of the full population.  It is not obvious that this equation will remain a prevalence-weighted sum when holding out samples, especially as the restricted-prevalence does not in general equal $p$.   Further implications of this observation are explored in the next section.

However, an immediate practical consequence of Eq.\ \eqref{eq:restrictedprev} is that {\it the relative fraction of positives from an assay using indeterminates is not a reliable estimator of total prevalence.}  In order for the restricted prevalence $p_\D$ to equal $p$, one requires
\begin{align}
\mathscr N_P - \mathscr N_Q = 0 = \int_{\D} P(\r) - pP(\r) - (1-p)N(\r) \d\r \nonumber, 
\end{align}
which implies
\begin{align}
0=\int_{\D}P(\r) - N(\r) = \mathscr N_P - \mathscr N_N.
\end{align}
That is, $p=p_\D$ only occurs when the holdout domain removes equal mass from the probability models, which is extremely restrictive.

To overcome this problem, we recall that Ref.\ (\cite{Patrone2021}), demonstrated how {\it an unbiased estimate of the total prevalence can be constructed without classifying samples} using a simple counting exercise on subdomains of $\Omega$.  The validity of that method is independent of the assay accuracy, so that it can be used to estimate $p$ in the present work.  Indeed, such techniques are necessary to construct the optimal classification domains, given the fundamental role of $p$ in their definitions.   We refer the reader to Ref.\ (\cite{Patrone2021}) for a deeper discussion of such issues.  

\subsection{Other Notions of Optimality}
\label{subsec:otheropt}

A common practice in the testing community is to preferentially optimize an assay so that either the specificity or sensitivity reaches a desired target, but not explicitly a linear combination of the two.  Equation \eqref{eq:noconundrum} and the bathtub principle suggest a route by which our method can solve an analogue of this problem.  However, a deeper investigation of sensitivity and specificity is first necessary to motivate this generalization and understand how such methods differ from Eqs.\ \eqref{eq:optindeterminate}--\eqref{eq:zdef}.  [See also Ref.\ (\cite{ROC}) for additional notions of optimality, as well as Refs.\ (\cite{3sig1,3sig2,3sig3}) for other approaches to defining classification domains.]  

Examination of the binary problem reveals that when $p=1/2$, the domains $D_P^\star$ and $D_N^\star$ equally weight sensitivity and specificity; that is, errors in either are treated as equally undesirable.  It is straightforward to show that increasing $p$ will increase sensitivity at the expensive of specificity, and vice versa.  The interpretation of this observation is that as the number of positive samples increases, we should increase the size of the positive classification domain so as to capture the their increasing share of the population.  It is therefore possible and even likely that when the prevalence approaches 0 or 100~\%, either sensitivity or specificity may be unacceptably low, since the corresponding contribution to the total accuracy becomes negligible.  

\begin{figure}
\includegraphics[width=8cm]{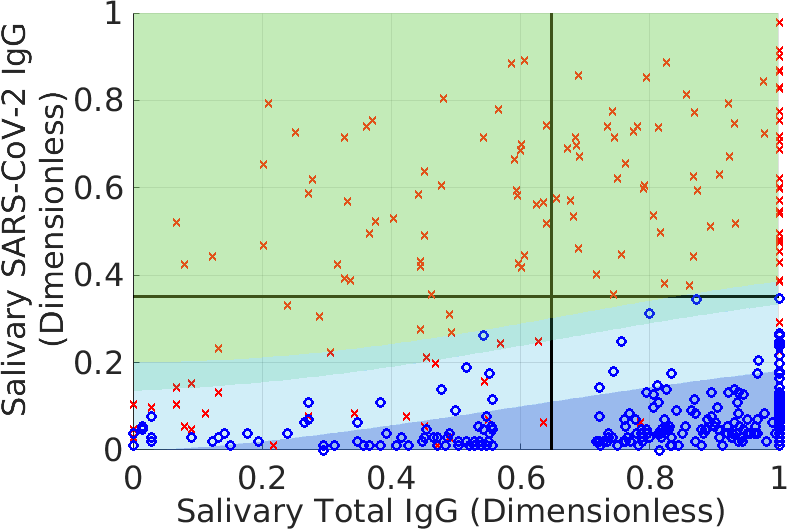}\caption{Holdout domain computed with a target accuracy of 99~\% and according to the constraints given by inequalities \eqref{eq:ineq2} and \eqref{eq:ineq3}.  For the latter, we set $X_-$ indirectly by holding out samples up to $Z_p=0.972$ in the positive classification domain.  This yields an empirical specificity of the training data was 100~\% while keeping the empirical sensitivity above 94~\%.  Note that the indeterminate domain (light-blue) is increased only into the positive classification domain (yellow-green) in attempting to satisfy inequality \eqref{eq:ineq3}.  The teal strip adjacent to the light blue and yellow-green is the modified indeterminate domain.  After increasing the empirical specificity to 100~\%, the optimized domains holds out $15.1~\%$ of samples, as opposed to $22.3~\%$ for the rectilinear method; see Table \ref{tab1}. }\label{fig:ineqhold}
\end{figure}

A possible solution to this problem is to recast Eq.\ \eqref{eq:constraint} as an inequality constraint of the form
\begin{align}
p\int_{\D_P} \d \r P(\r) + (1-p)\int_{\D_N} \d \r N(\r) \ge X \int_{\D} Q(\r) \label{eq:ineq1}
\end{align}
together with the additional constraints
\begin{align}
S_e \ge X_+  \label{eq:ineq2}\\ 
S_p \ge X_-, \label{eq:ineq3}
\end{align}
where $X_+$ and $X_-$ are user-defined lower bounds.  While an optimal solution to this problem is beyond the scope of the current manuscript, the bathtub principle suggests a construction akin to active-set methods (\cite{Nocedal}).  First, solve the optimization problem associated with Eqs.\ \eqref{eq:objective}--\eqref{eq:constraint} and check the resulting values of sensitivity and specificity.  If these quantities are deemed to small, remove samples up to user-defined waterlines $Z_n\ge Z_0$ and $Z_p\ge Z_0$ (which may be different), where $Z_n$ and $Z_p$ apply only to samples in the negative and positive classification domains.  Figure \ref{fig:ineqhold} shows an example of this approach applied to the data in previous figures.  We originally set $X=0.99$ but required that the empirical specificity be 100~\% for the training set.  To accomplish this, we set $Z_p = 0.972$, which augments the size of the indeterminate domain (teal strip added to the light blue domain) without decreasing the number of true negatives.


\subsection{Relationship between Prevalence, Sensitivity, and Specificity}

Equation \eqref{eq:noconundrum} and the examples of Secs.\ \eqref{subsec:prev} and \eqref{subsec:otheropt} beg the question: to what extent is  prevalence-weighted accuracy a preferred or natural framework for diagnostic classification, as opposed to methods based on explicit reference to sensitivity and specificity?  To unravel this, consider that the latter two are purely theoretical properties of a specific choice of classification domain and are only loosely connected to the reality of testing.  This is evident from the definitions given by Eqs.\ \eqref{eq:sensitivity} and \eqref{eq:specificity}.  The concept of prevalence, {\it i.e.\ implying existence of a population}, does not enter; rather all that is needed is a choice of the classification domains.  Thus, an assay can have exceptional sensitivity and yet still be wrong half the time if the prevalence is low.  In a related vein, it is clear that specificity and sensitivity only characterize assay accuracy in the limits $p\to 0$ and $p\to 1$, respectively.

Here we encourage a new perspective.  As a baseline strategy, the most important task is to correctly classify samples; at least this is of the utmost importance to patients.  Moreover, computing accurate prevalence estimates is critical for epidemiologists (although we have shown previously that this problem is solved accurately without recourse to classification).  With this goal in mind, the sensitivity and specificity are subservient to accuracy via Eq.\ \eqref{eq:constraint}, and it is not unreasonable to let them change with prevalence if doing so increases overall testing accuracy.  We highlight this because under such a paradigm, $S_e$ and $S_p$ lose their status as the key performance metrics that define the ``quality'' of an assay, and they cannot be viewed as static properties.  Such observations are not to say that $S_e$ and $S_p$ are useless, however.   Clearly there are times when it is more important to correctly identify samples from one class, and this motivates the generalization of Sec.\ \ref{subsec:otheropt}.  

But these observations clarify our perspective of why the prevalence sets a natural scale for classification.  In particular, Eq.\ \eqref{eq:constraint} has two equivalent interpretations: (i) the accuracy of the assay must be $X$; and (ii) the prevalence-weighted sensitivity and specificity must be $X$.  The equivalence of these interpretations arises from the fact that {\it notions of accuracy assume the existence of a population to which the test is applied.}  Thus, Eq.\ \eqref{eq:noconundrum} is perhaps unsurprising in light of Eq.\ \eqref{eq:constraint} because both are self-consistent statements about the properties of a population.

The benefit of treating prevalence-weighting as a natural framework for diagnostic classification is that one can easily identify when subjective elements (i.e.\ not intrinsic to the population) have been added to the analysis.  For example, the indeterminate domain in Fig.\ \ref{fig:ineqhold} associated with the inequalities \eqref{eq:ineq1} -- \eqref{eq:ineq3} is not optimal insofar as there is a smaller counterpart that yields the same average accuracy for the classified data.  However, it is clear by construction how we have modified the latter, i.e.\ by adding a user-defined constraint on the specificity.  Likewise, even Eq.\ \eqref{eq:constraint} should be viewed as a subjective modification of the unconstrained, prevalence-weighted classification problem.  

Ultimately the choice of classification method is best determined by assay developers, and there may be situations in which prevalence weighting is inappropriate.  Nonetheless, we feel that the analysis herein highlights the assumptions behind our work and attempts to ground it in objective elements inherent to the population of interest.

\subsection{Limitations and Open Directions}

A fundamental limitation of our analysis is the assumption that the probabilistic models describing positive and negative samples can be used outside the scope of training data.  This problem is common to virtually any classification scheme and is primarily an issue of modeling.  Such issues have been explored in a previous manuscript, to which we refer the reader (\cite{Patrone2021}).  We note here, however, that model-form errors may introduce uncertainty on the order of a few percent in the conditional probability densities.  Thus, it is likely that modeled estimates of accuracy will be incorrect by a proportional amount.  This is seen, for example, in the holdout domain computed in Fig.\ \ref{fig:holdout}.  However, Sec.\ \ref{subsec:otheropt} provides means of ensuring that the indeterminate domains are recomputed to satisfy any constraints on empirical estimates of accuracy.  We also note that approaches that do not explicitly account for prevalence and/or conditional probabilities are likely to have significantly more model-form errors than estimates based on our approach.

Regarding the indeterminate analysis, Eqs.\ \eqref{eq:optindeterminate}--\eqref{eq:zdef} and the generalization considered in Sec.\ \ref{subsec:otheropt} may be a challenging optimization problem to solve, although the solution could be extremely useful for satisfying regulatory and/or public health requirements.  Moreover, formalizing the algorithm described in that section and studying its properties relative to the optimal solution may be useful.

A practical limitation of our analysis is the definition of assay performance, provided we allow for variable, prevalence-dependent classification domains.  Current standards advocate using sensitivity and specificity estimated for a single validation population having a fixed prevalence.  To realize the full potential of our analysis, it is necessary to (i) estimate assay accuracy and uncertainty therein, (ii) characterize the admissible classification domains, and (iii) compute sensitivities and specificities, all as a function of the variable prevalence.  While such issues have been partly considered in (\cite{Patrone2021}), and deeper investigation of this uncertainty quantification is necessary for widespread adoption of these techniques.  

{\bf Acknowledgements:}This work is a contribution of the National Institute of Standards and Technology and is not subject to copyright in the United States.  Funding for Johns Hopkins University authors was provided by the Johns Hopkins COVID-19 Research and Response Program, the FIA Foundation, a gift from the GRACE Communications Foundation (C.D.H., N.P.), National Cancer Institute (NCI) SeroNet grant U01CA260469 (C.D.H.), National Institute of Allergy and Infectious Diseases (NIAID) grant R21AI139784 (C.D.H. and N.P), National Institute of Environmental Health Sciences (NIEHS) grant R01ES026973 (C.D.H., N.P.), NIAID grant R01AI130066 and NIH grant U24OD023382 (C.D.H), NIAID grant 3R01AI148049 (D.L.T.), the Johns Hopkins University School of Medicine COVID-19 Research Fund, the Sherrilyn and Ken Fisher Center for Environmental Infectious Diseases Discovery Program, and NIH grants U54EB007958-12, U5411090366 (Y.C.M.).  The aforementioned funders had no role in study design, data analysis, decision to publish, or preparation of the manuscript.  P.B. was also funded through the NIST PREP grant 70NANB18H162.

{\bf Research involving Human Participants and/or Animals:} Use of data provided in this manuscript has been approved by: (1) the NIST Research Protections Office; and (2) the Johns Hopkins School of Medicine Internal Review Board.

{\bf Data Availability:}  Analysis scripts and data developed as a part of this work are available upon reasonable request.  
\appendix

\section{Proof of main result}
\label{sec:proof}

{\bf Lemma 1}: {\it Assume that $P(\r)$ and $N(\r)$ are summable functions on $\Omega$ and that the measure of any point $\r$ is zero with respect to all distributions.  Also assume that $\L[D_P^\star,D_N^\star]< X$ and that there exists a set of non-zero measure for which $Z(\r) > X$. Then the sets defined by Eqs.\ \eqref{eq:optindeterminate}--\eqref{eq:zdef} minimize Eq.\ \eqref{eq:objective} subject to Eq.\ \eqref{eq:constraint}. }

We first show that Eq.\ \eqref{eq:zdef} defines $Z_0(X)$ \textcolor{black}{and $\mathcal C(X)$.}  Let $1/2 \le \zeta \le 1$ and define
\begin{subequations}
\begin{align}
\D_h(\zeta) &= \{\r: Z^\star(\r) < \zeta\} \\
\D_P(\zeta) &=  D_P^\star / \D_h(\zeta)\\
\D_N(\zeta) &= D_N^\star / \D_h(\zeta).
\end{align}
\end{subequations}
Equation \eqref{eq:zdef} motivates the function
\begin{align}
I(\zeta) =\left[\int_{\D_P(\zeta) \cup \D_N(\zeta)} \hspace{-10mm} \d\r\, Q(\r)\right]^{-1} \left[\int_{\D_P(\zeta) \cup \D_N(\zeta)} \hspace{-15mm} \d\r\, [Z^\star(\r) - X]Q(\r)\right],
\end{align}
which is a monotone increasing function of $\zeta$ satisfying the inequalities $I(1/2) < 0$ and $I(\zeta) > 0$ for some $\zeta >1/2$.  Thus, there exists a unique value of $Z_0(X)$ for which one of two situations holds: either (I) the function $I(\zeta)$ is continuous at $Z_0(X)$ and  $I(Z_0(X)) = 0$, which directly implies Eq.\ \eqref{eq:zdef}; or (II) $I(\zeta)$ suffers a discontinuity, so that $I(Z_0(X)) < 0$ and $I(Z_0(X) + \epsilon) > 0$ for any positive $\epsilon$.  The latter case occurs when $\mathcal S = \{\r: Z^\star(\r) = Z_0(X)\}$ has non-zero measure, and we may set $\mathcal C$ to be any subset $\mathcal C \subset \mathcal S$ provided Eq.\ \eqref{eq:zdef} is satisfied.  The existence of such a $\mathcal C$ is guaranteed by the linearity of integration, which implies that 
 \begin{align}
\hat I(C) =\left[\int_{\D_P(Z_0) \cup \D_N(Z_0) \cup C} \hspace{-10mm} \d\r\, Q(\r)\right]^{-1} \left[\int_{\D_P(Z_0) \cup \D_N(Z_0) \cup C} \hspace{-15mm} \d\r\, [Z^\star(\r) - X]Q(\r)\right] \nonumber
\end{align}
is a continuous, monotone increasing function of the measure of $C \subset \mathcal S$ that passes through zero.  Any zero of $\hat I(C)$ implies Eq.\ \eqref{eq:zdef} and defines an appropriate $\mathcal C$.

The proof that Eqs.\ \eqref{eq:optindeterminate}--\eqref{eq:zdef} minimize Eq.\ \eqref{eq:objective} relies on the observation that any $Z^\star(\r) < Z_0(X)$ is farther from the mean value $X$ than any $Z^\star(\r) > Z_0(X)$.  Thus, it ``costs additional probability'' to swap points between the indeterminate region and $\D^\star=\D_P^\star\cup \D_N^\star$ while satisfying the constraint.  To see this mathematically, let $\D$ be any other union of positive and negative classification domains satisfying Eq.\ \eqref{eq:constraint}.  We do not consider any domains $\D$ that consist only of choosing a different subset $\mathcal C \subset \mathcal S$ while maintaining Eq.\ \eqref{eq:zdef}.  By Eq.\ \eqref{eq:constraint} one find
\begin{align}
\int_{\D/\D^\star}\!\!\! \d \r\, Q(\r) [X - Z^\star(\r) ] - \int_{\D^\star / \D}\!\!\! \d \r\, Q(\r)[X - Z^\star(\r)] = 0 \label{eq:lookconstraint}
\end{align}
We can further expand the second term as
\begin{align}
&\int_{\D^\star /\D}\hspace{-6mm}\d\r\,\, Q(\r) [X- Z^\star(\r)] \nonumber \\
&\qquad =\int_{\Z^+ / \D}\hspace{-6mm} \d \r\, Q(\r)[X- Z^\star(\r) ] + \int_{\Z^-/\D}\hspace{-6mm} \d \r\, Q(\r) [X- Z^\star(\r)] \label{eq:domainsep}
\end{align}
where $\Z^+=\{r:Z^\star(\r) > X \}$ and $\Z^- = \{r:Z_0(X)<Z^\star(\r) < X\}$.  Clearly the first term on the RHS of Eq.\ \eqref{eq:domainsep} is negative, whereas the second term is positive.  Noting that $Z^\star(\r \in \D^\star) > Z^\star(\r \in \D_h^\star)$, one finds by inserting Eq.\ \eqref{eq:domainsep} into Eq.\ \eqref{eq:lookconstraint} that the latter can be expressed in the form
\begin{align}
\int_{\D/\D^\star}\hspace{-6mm}\d\r\, Q(\r)A(\r) = \int_{\Z^+ / \D}\hspace{-6mm} \d \r\, Q(\r)B(\r) + \int_{\Z^-/\D}\hspace{-6mm} \d \r\, Q(\r)C(\r)
\end{align}
where $A(\r) > 0$, $B(\r) < 0$, and $0 < C(\r) < A(\r)$.  This implies that
\begin{align}
\int_{\D/\D^\star}\hspace{-6mm}\d\r\, Q(\r) < \int_{\D^\star / \D}\hspace{-6mm} \d \r\, Q(\r) \implies \int_{\D_h/\D_h^\star}\hspace{-6mm}\d\r\, Q(\r) > \int_{\D_h^\star / \D_h}\hspace{-6mm} \d \r\, Q(\r). \label{eq:probinequality}
\end{align}
Consider now the difference of objective functions
\begin{align}
\Delta \H = \H[\D_h] - \H[\D_h^\star] &= \int_{\D_h} \hspace{-3mm}\d\r\,Q(\r) - \int_{\D_h^\star}\hspace{-3mm}\d\r\,Q(\r) \nonumber \\
&= \int_{\D_h/\D_h^\star} \hspace{-6mm}\d\r\,Q(\r) - \int_{\D_h^\star/\D_h}\hspace{-6mm}\d\r\,Q(\r).
\end{align}
By inequality \eqref{eq:probinequality}, we see that $\Delta \H > 0$.  Moreover, note that $Z(\r,D_P,D_N) \le Z^\star(\r)$ for any classification domains associated with the binary problem.  Clearly any choice besides $D_N^\star$ and $D_P^\star$ entails increasing the measure of $\D_h$ to ensure that the constraint is satisfied. \qed

\section{On PDFs with Dirac Masses}

Figure \ref{fig:rawdata} illustrates that biological phenomena may generate a signal so strong that the instrument saturates, i.e.\ it reaches a limit $x_{\max}$ above which it cannot distinguish different measurement values.  This saturation effectively rounds the ``true'' measurement down to the $x_{\max}$.  The only conclusion we can draw about a reported value $x_{\max}$ is that the true value $\chi$ satisfies the inequality $\chi \ge x_{\max}$.  Similar there exists a lower limit $x_{\min}$ up to which smaller measurements values are rounded.  The goal of this section is to incorporate such information into probability modeling.

For concreteness, we restrict ourselves to the one dimensional measurements $x$ associated with the total IgG assay.  We assume that were the optical photodetector not restricted to the range $[x_{\min},x_{\max}]$, the recorded measurement would have been $\chi$ returned on the domain $-\infty < \chi < \infty$.   Because the measurements have been transformed to a logarithmic coordinate system, $\chi \to - \infty$ is meaningful.  Without additional information about probability of total IgG antibody levels, we make a minimal assumption that $\chi$ is described by a Gaussian distribution with an unknown mean $\mu$ and variance $\sigma^2$.  Thus, on the {\it open} domain $(x_{\min},x_{\max})$, assume that $x=\chi$, so that the probability of measuring $x$ is
\begin{align}
\hat P_0(x) = \frac{1}{\sqrt{2\pi}\sigma}e^{-\frac{(x-\mu)^2}{2\sigma^2}}, \hspace{10mm} x_{\min} < x < x_{\max}. \label{eq:Gaussianmodel}
\end{align}
However, on the boundaries $x_{\min}$ and $x_{\max}$, we only know that the true values are below and above the respective thresholds.  Thus, the probabilities of measuring  $x_{\min}$ and $x_{\max}$ are given by
\begin{align}
\hat P_l &= \int_{-\infty}^{x_{\min}} \hat P_0(\chi|\mu,\sigma^2)\, \d \chi \\
\hat P_r &= \int_{x_{\max}}^{\infty} \hat P_0(\chi|\mu,\sigma^2)\, \d \chi
\end{align} 
where $\hat P_0(\chi|\mu,\sigma^2)$ is the same as Eq.\ \eqref{eq:Gaussianmodel}, but with $x$ replaced by $\chi$.  We may then write the full probability model for $x$ as
\begin{align}
\hat P(x) &= \frac{1}{\sqrt{2\pi}\sigma}e^{-\frac{(x-\mu)^2}{2\sigma^2}} \mathbb I(x,x_{\min},x_{\max}) \nonumber \\ 
& \qquad \qquad + \delta(x-x_{min})\hat P_l + \delta(x-x_{\max}) \hat P_r. \label{eq:fullprobmodel}
\end{align}
where $\mathbb I(x,a,b)$ is the indicator function that $x$ is in the open set $(a,b)$.

Equation \eqref{eq:fullprobmodel} motivates a generalization of the MLE.  Define the likelihood function 
\begin{align}
L(x) = 
\begin{cases}
\hat P_0(x) & x_{\min} < x < x_{\max} \\
\hat P_l & x = x_{\min} \\
\hat P_r & x = x_{\max}
\end{cases}.
\end{align}
To determine the values of $\sigma$ and $\mu$, we maximize with respect to these parameters the product of $N$ likelihoods given by
\begin{align}
\L_{\rm like}(\boldsymbol {\rm x}) = \prod_{j=1}^N L(x_j),
\end{align}
or alternatively, we minimize the negative log of $\L_{\rm like}(\boldsymbol {\rm x})$.  To construct the two-dimensional PDF associated with Eq.\ \eqref{eq:formalPDF}, we assume the corresponding probability model for the SARS-CoV-2 IgG measurements and use standard MLE to identify the distribution parameters.  The full PDF for training data is then given by the product of the corresponding PDFs for total IgG and SARS-CoV-2 measurements and has the form given by Eq.\ \eqref{eq:formalPDF}.

Note that Eq.\ \eqref{eq:formalPDF} does not require modification of the proof in the previous section, since any point $(x,y)$ is a set of measure zero, provided that $\mathcal P_l(y)$ and $\mathcal P_r(y)$ (and their negative counterparts) are bounded functions of $y$. However, we do require care in defining the local accuracy and classification domains $D^\star_N$ and $D_P^\star$.  Let
\begin{subequations}
\begin{align}
A_+ &= \{\r: p\mathcal P_0(\r) > (1-p)\mathcal N_0(\r), x_{\min} < x < x_{\max}\} \\
B_+ &= \{\r: p\mathcal P_l(y) > (1-p)\mathcal N_l(y), x= x_{\min}\} \\
C_+ &=\{\r: p\mathcal P_r(y) > (1-p)\mathcal N_r(y), x= x_{\max}\}
\end{align}
\end{subequations}
from which we construct $D_P^\star = A_+ \cup B_+ \cup C_+$ and the analogous definition for $D_N^\star$.  Note that any point for which the prevalence-weighted probabilities of being negative and positive are identical can be assigned to either class.  The corresponding definition of $Z^\star(\r)$ is given by
\begin{align}
Z^\star(\r) = \begin{cases}
\frac{\max[p\mathcal P_0(\r),(1-p)\mathcal N_0(\r)]}{p\mathcal P_0(\r)+(1-p)\mathcal N_0(\r)} & x_{\min} < x < x_{\max} \\
\frac{\max[p \mathcal P_l(y),(1-p) \mathcal N_l(y)]}{p \mathcal P_l(y)+(1-p) \mathcal N_l(y)} &  x = x_{\min} \\
\frac{\max[p \mathcal P_r(y),(1-p) \mathcal N_r(y)]}{p \mathcal P_r(y)+(1-p) \mathcal N_r(y)} &  x = x_{\max} 
\end{cases}
\end{align}
where $\N_l(y)$ and $\N_r(y)$ are the analogous of $\P_l(y)$ and $\P_r(y)$ for the negative PDF.  

\section{On the Point-Swap Derivatives}

To justify the use of Eq.\ \eqref{eq:setderivative}, return to Eq.\ \eqref{eq:constraint} and consider a set $\D$ and its complement $\D_h$.  Consider balls $\mathcal B=B(\r,\epsilon)$ and $\mathcal B'=B(\r',\epsilon')$ having radii $\epsilon$, $\epsilon'$ and centered about $\r$ and $\r'$.  Let these balls be entirely contained in $\D_h$ and $\D$, respectively.  Momentarily assume that the PDFs do not contain Dirac masses.  Define $\D_h'$ and $\D'$ to be the sets where $\mathcal B$ and $\mathcal B'$ have been interchanged without violating Eq.\ \eqref{eq:constraint}.  Taking the difference of Eq.\ \eqref{eq:constraint} defined relative to $\D$ and $\D'$ yields
\begin{align}
\int_{\mathcal B}\d \hat \r [Z(\hat \r)-X]Q(\hat \r) - \int_{\mathcal B'} \d\hat \r [Z(\hat \r)-X]Q(\hat \r)=0.
\end{align}
Assuming that $Z(\hat \r)$ and $Q(\hat \r)$ are sufficiently smooth, to leading order in $\epsilon$, $\epsilon'$ one finds
\begin{align}
\epsilon^d [Z(\r) - X]Q(\r) = (\epsilon')^d [Z(\r') - X]Q(\r') .
\end{align}
where $d$ is the dimensionality of $\r$. Rearranging this last equation yields
\begin{align}
\frac{(\epsilon')^d Q(\r')}{\epsilon^d Q(\r)} = \frac{Z(\r) - X}{Z(\r') - X}. \label{eq:swapdef}
\end{align}
Note that $\epsilon^d$ and $(\epsilon')^d$ a proportional to the volumes of the respective balls about the points $\r$ and $\r'$, so that the quantity $(\epsilon')^d Q(\r')$ is, for example, the infinitesimal probability mass contained in the corresponding ball.  Thus, the given by Eq.\ \eqref{eq:swapdef} is the relative change probability mass exchanged between $\D$ and $\D_h$ in swapping $\r$ and $\r'$.

If we change the class of $\r$ (either from $\D_P$ to $\D_N$ or vice versa), it may be necessary to hold out additional points $\r'$, or it may be possible to move points from the indeterminate into the classification domain.  In either case, letting $\mathcal B$ and $\mathcal B'$ have the same definitions as before and assuming Eq.\ \eqref{eq:constraint} holds, one finds
\begin{align}
\int_{\B}\d\hr [2Z(\hr) - 1]Q(\hr)  \pm \int_{\B'}\d\hr [ Z(\hr) - X]Q(\hr) = 0, \label{eq:class_swap}
\end{align}
where $\B'$ is the ball moved to ($+$) or from ($-$) the indeterminate domain, depending on the sign of the first term; note that we also require $Z < X$ inside $\B'$.  Again taking the limit that the respective $\epsilon$ are small, one finds
\begin{align}
\epsilon^d[2Z(\r) - 1]Q(\r)  = \pm (\epsilon')^d[X-Z(\r')]Q(\r')
\end{align}
Dividing by $\pm [X-Z(\r')]$ yields the infinitesimal probability mass moved to or from the indeterminate
\begin{align}
(\epsilon')^d Q(\r') =  \pm\frac{\epsilon^d[2Z(\r) - 1]Q(\r)}{X-Z(\r')}.
\end{align}
The LHS must be positive, and the denominator on the RHS is positive.  Thus, the $+$ and $-$ signs on the RHS occur when $Z(\r) > 1/2$ and $Z(\r) < 1/2$, corresponding to the situations in which probability moves to and from the indeterminate region.  Thus, in assessing when $\D_h$ grows, it is sufficient to test the inequality $Z(\r) > 1/2$.
 
The analysis of this section is easily generalized to the case of Eq.\ \eqref{eq:formalPDF} by noting that for points on the lines $x=0$ and $x=1$, the balls of radius $\epsilon$ should be taken as intervals on the line with length $2\epsilon$.  This yields the appropriate generalization of probability associated with those points.
 
 \bibliographystyle{agsm}

\bibliography{holdout}

@article{Heaney21,
title = {Comparative performance of multiplex salivary and commercially available serologic assays to detect SARS-CoV-2 IgG and neutralization titers},
journal = {Journal of Clinical Virology},
volume = {145},
pages = {104997},
year = {2021},
issn = {1386-6532},
author = {Christopher D. Heaney and Nora Pisanic and Pranay R. Randad and Kate Kruczynski and Tyrone Howard and Xianming Zhu and Kirsten Littlefield and Eshan U. Patel and Ruchee Shrestha and Oliver Laeyendecker and Shmuel Shoham and David Sullivan and Kelly Gebo and Daniel Hanley and Andrew D. Redd and Thomas C. Quinn and Arturo Casadevall and Jonathan M. Zenilman and Andrew Pekosz and Evan M. Bloch and Aaron A.R. Tobian}
}

@book{Nocedal,
  title={Numerical Optimization},
  author={Nocedal, J. and Wright, S.},
  isbn={9780387400655},
  lccn={2006923897},
  series={Springer Series in Operations Research and Financial Engineering},
  year={2006},
  publisher={Springer New York}
}

@Article{ROC,
author={Florkowski, Christopher M.},
title={Sensitivity, specificity, receiver-operating characteristic (ROC) curves and likelihood ratios: communicating the performance of diagnostic tests},
journal={The Clinical biochemist. Reviews},
year={2008},
month={Aug},
publisher={The Australian Association of Clinical Biochemists},
volume={29 Suppl 1},
number={Suppl 1},
pages={S83-S87}
}

@article{Nora1,
author = {Nora Pisanic  and Pranay R. Randad  and Kate Kruczynski  and Yukari C. Manabe  and David L. Thomas  and Andrew Pekosz  and Sabra L. Klein  and Michael J. Betenbaugh  and William A. Clarke  and Oliver Laeyendecker  and Patrizio P. Caturegli  and H. Benjamin Larman  and Barbara Detrick  and Jessica K. Fairley  and Amy C. Sherman  and Nadine Rouphael  and Srilatha Edupuganti  and Douglas A. Granger  and Steve W. Granger  and Matthew H. Collins  and Christopher D. Heaney  and Michael J. Loeffelholz },
title = {COVID-19 Serology at Population Scale: SARS-CoV-2-Specific Antibody Responses in Saliva},
journal = {Journal of Clinical Microbiology},
volume = {59},
number = {1},
pages = {e02204-20},
year = {2020},
doi = {10.1128/JCM.02204-20}
}

@article {Nora2,
	author = {Randad, Pranay R. and Pisanic, Nora and Kruczynski, Kate and Howard, Tyrone and Rivera, Magdielis Gregory and Spicer, Kristoffer and Antar, Annukka A.R. and Penson, Tristan and Thomas, David L. and Pekosz, Andrew and Ndahiro, Nelson and Aliyu, Lateef and Betenbaugh, Michael J. and Manley, Hannah and Detrick, Barbara and Katz, Morgan and Cosgrove, Sara and Rock, Clare and Zyskind, Israel and Silverberg, Jonathan I. and Rosenberg, Avi Z. and Duggal, Priya and Manabe, Yukari C. and Collins, Matthew H. and Heaney, Christopher D.},
	title = {Durability of SARS-CoV-2-specific IgG responses in saliva for up to 8 months after infection},
	elocation-id = {2021.03.12.21252149},
	


year = {2021}
}

@article{Patrone2021,
    author = {Patrone, Paul N and Kearsley, Anthony J},
    title = "{Classification under uncertainty: data analysis for diagnostic antibody testing}",
    journal = {Mathematical Medicine and Biology: A Journal of the IMA},
    year = {2021},
    month = {08},
    issn = {1477-8602},
    doi = {10.1093/imammb/dqab007},
    note = {dqab007}
}

@article {challenges2,
	author = {Bermingham, William H and Wilding, Thomas and Beck, Sarah and Huissoon, Aarnoud},
	title = {SARS-CoV-2 serology: Test, test, test, but interpret with caution!},
	volume = {20},
	number = {4},
	pages = {365--368},
	year = {2020},
	doi = {10.7861/clinmed.2020-0170},
	publisher = {Royal College of Physicians},
	journal = {Clinical Medicine}
}

@article{geography,
title = "The COVID-19 Serology Studies Workshop: Recommendations and Challenges",
journal = "Immunity",
volume = "53",
number = "1",
pages = "1 - 5",
year = "2020",
issn = "1074-7613",
author = "Andrea M. Lerner and Robert W. Eisinger and Douglas R. Lowy and Lyle R. Petersen and Rosemary Humes and Matthew Hepburn and M. Cristina Cassetti"
}

@article{challenges1,
    author = {Bond, Katherine and Nicholson, Suellen and Lim, Seok Ming and Karapanagiotidis, Theo and Williams, Eloise and Johnson, Douglas and Hoang, Tuyet and Sia, Cheryll and Purcell, Damian and Mordant, Francesca and Lewin, Sharon R and Catton, Mike and Subbarao, Kanta and Howden, Benjamin P and Williamson, Deborah A},
    title = "{Evaluation of Serological Tests for SARS-CoV-2: Implications for Serology Testing in a Low-Prevalence Setting}",
    journal = {The Journal of Infectious Diseases},
    volume = {222},
    number = {8},
    pages = {1280-1288},
    year = {2020},
    month = {08},
    doi = {10.1093/infdis/jiaa467}
}

@misc{EUA,
  author={FDA},
  year={2020},
  title = {EUA Authorized Serology Test Performance},
  howpublished = {{https://www.fda.gov/medical-devices/coronavirus-disease-2019-covid-19-emergency-use-authorizations-medical-devices/eua-authorized-serology-test-performance}},
  note = {Accessed: 2020-09-16}
}

@Article{holdout1,
author={Meyer, B.
and Torriani, G.
and Yerly, S.
and Mazza, L.
and Calame, A.
and Arm-Vernez, I.
and Zimmer, G.
and Agoritsas, T.
and Stirnemann, J.
and Spechbach, H.
and Guessous, I.
and Stringhini, S.
and Pugin, J.
and Roux-Lombard, P.
and Fontao, L.
and Siegrist, C.-A.
and Eckerle, I.
and Vuilleumier, N.
and Kaiser, L.
and for Emerging Viral Diseases, Geneva Center},
title={Validation of a commercially available SARS-CoV-2 serological immunoassay},
journal={Clinical microbiology and infection : the official publication of the European Society of Clinical Microbiology and Infectious Diseases},
year={2020},
month={Oct},
edition={2020/06/27},
publisher={European Society of Clinical Microbiology and Infectious Diseases. Published by Elsevier Ltd.},
volume={26},
number={10},
pages={1386-1394},
note={32603801[pmid]},
note={PMC7320699[pmcid]},
note={S1198-743X(20)30368-2[PII]},
issn={1469-0691},
doi={10.1016/j.cmi.2020.06.024},
language={eng}
}

@article{holdout2,
    doi = {10.1371/journal.pone.0253889},
    author = {Lee, Nuri AND Jeong, Seri AND Park, Min-Jeong AND Song, Wonkeun},
    journal = {PLOS ONE},
    publisher = {Public Library of Science},
    title = {Comparison of three serological chemiluminescence immunoassays for SARS-CoV-2, and clinical significance of antibody index with disease severity},
    year = {2021},
    month = {06},
    volume = {16},
    pages = {1-13},
    number = {6}
}

@article{holdout3,
    author = {Manthei, David M and Whalen, Jason F and Schroeder, Lee F and Sinay, Anthony M and Li, Shih-Hon and Valdez, Riccardo and Giacherio, Donald A and Gherasim, Carmen},
    title = "{Differences in Performance Characteristics Among Four High-Throughput Assays for the Detection of Antibodies Against SARS-CoV-2 Using a Common Set of Patient Samples}",
    journal = {American Journal of Clinical Pathology},
    volume = {155},
    number = {2},
    pages = {267-279},
    year = {2020},
    month = {10},
    issn = {0002-9173},
    doi = {10.1093/ajcp/aqaa200}
}

@Article{holdout4,
author={Theel, Elitza S.
and Harring, Julie
and Hilgart, Heather
and Granger, Dane},
title={Performance Characteristics of Four High-Throughput Immunoassays for Detection of IgG Antibodies against SARS-CoV-2},
journal={Journal of clinical microbiology},
year={2020},
month={Jul},
day={23},
publisher={American Society for Microbiology},
volume={58},
number={8},
pages={e01243-20},
note={32513859[pmid]},
note={PMC7383546[pmcid]},
note={JCM.01243-20[PII]},
issn={1098-660X},
doi={10.1128/JCM.01243-20},
language={eng}
}

@book{bathtub,
  title={Analysis},
  author={Lieb, E.H. and Loss, M. and LOSS, M.A. and American Mathematical Society},
  isbn={9780821827833},
  lccn={01018215},
  series={Crm Proceedings \& Lecture Notes},
  year={2001},
  publisher={American Mathematical Society}
}

@article{Chou21,
author = {Böttcher, Lucas  and D'Orsogna, Maria R.  and Chou, Tom },
title = {A statistical model of COVID-19 testing in populations: effects of sampling bias and testing errors},
journal = {Philosophical Transactions of the Royal Society A: Mathematical, Physical and Engineering Sciences},
volume = {380},
number = {2214},
pages = {20210121},
year = {2022},
doi = {10.1098/rsta.2021.0121}
}

@Article{3sig3,
author={Hachim, Asmaa
and Kavian, Niloufar
and Cohen, Carolyn A.
and Chin, Alex W. H.
and Chu, Daniel K. W.
and Mok, Chris K. P.
and Tsang, Owen T. Y.
and Yeung, Yiu Cheong
and Perera, Ranawaka A. P. M.
and Poon, Leo L. M.
and Peiris, J. S. Malik
and Valkenburg, Sophie A.},
title={ORF8 and ORF3b antibodies are accurate serological markers of early and late SARS-CoV-2 infection},
journal={Nature Immunology},
year={2020},
month={Oct},
day={01},
volume={21},
number={10},
pages={1293-1301},
doi={10.1038/s41590-020-0773-7}
}

@article {3sig2,
	author = {Grzelak, Ludivine and Temmam, Sarah and Planchais, Cyril and Demeret, Caroline and Tondeur, Laura and Huon, Christ{\`e}le and Guivel-Benhassine, Florence and Staropoli, Isabelle and Chazal, Maxime and Dufloo, Jeremy and Planas, Delphine and Buchrieser, Julian and Rajah, Maaran Michael and Robinot, Remy and Porrot, Fran{\c c}oise and Albert, M{\'e}lanie and Chen, Kuang-Yu and Crescenzo-Chaigne, Bernadette and Donati, Flora and Anna, Fran{\c c}ois and Souque, Philippe and Gransagne, Marion and Bellalou, Jacques and Nowakowski, Mireille and Backovic, Marija and Bouadma, Lila and Le Fevre, Lucie and Le Hingrat, Quentin and Descamps, Diane and Pourbaix, Annabelle and Laou{\'e}nan, C{\'e}dric and Ghosn, Jade and Yazdanpanah, Yazdan and Besombes, Camille and Jolly, Nathalie and Pellerin-Fernandes, Sandrine and Cheny, Olivia and Ungeheuer, Marie-No{\"e}lle and Mellon, Guillaume and Morel, Pascal and Rolland, Simon and Rey, Felix A. and Behillil, Sylvie and Enouf, Vincent and Lemaitre, Audrey and Cr{\'e}ach, Marie-Aude and Petres, Stephane and Escriou, Nicolas and Charneau, Pierre and Fontanet, Arnaud and Hoen, Bruno and Bruel, Timoth{\'e}e and Eloit, Marc and Mouquet, Hugo and Schwartz, Olivier and van der Werf, Sylvie},
	title = {A comparison of four serological assays for detecting anti{\textendash}SARS-CoV-2 antibodies in human serum samples from different populations},
	volume = {12},
	number = {559},
	elocation-id = {eabc3103},
	year = {2020},
	doi = {10.1126/scitranslmed.abc3103},
	publisher = {American Association for the Advancement of Science}
}

@Article{3sig1,
author={Algaissi, Abdullah
and Alfaleh, Mohamed A.
and Hala, Sharif
and Abujamel, Turki S.
and Alamri, Sawsan S.
and Almahboub, Sarah A.
and Alluhaybi, Khalid A.
and Hobani, Haya I.
and Alsulaiman, Reem M.
and AlHarbi, Rahaf H.
and ElAssouli, M.-Z.aki
and Alhabbab, Rowa Y.
and AlSaieedi, Ahdab A.
and Abdulaal, Wesam H.
and Al-Somali, Afrah A.
and Alofi, Fadwa S.
and Khogeer, Asim A.
and Alkayyal, Almohanad A.
and Mahmoud, Ahmad Bakur
and Almontashiri, Naif A. M.
and Pain, Arnab
and Hashem, Anwar M.},
title={SARS-CoV-2 S1 and N-based serological assays reveal rapid seroconversion and induction of specific antibody response in COVID-19 patients},
journal={Scientific Reports},
year={2020},
month={Oct},
day={06},
volume={10},
number={1},
pages={16561},
doi={10.1038/s41598-020-73491-5}
}

\end{document}